\documentclass[runningheads]{llncs}
\usepackage[T1]{fontenc}

\usepackage{graphicx}
\usepackage{color}
\usepackage{amsmath,amssymb,amsfonts}
\usepackage{pifont}
\usepackage{multirow}
\usepackage{makecell}
\usepackage{bm}
\usepackage{booktabs}
\usepackage[misc]{ifsym}
\usepackage{algorithm}
\usepackage{algpseudocode}

\definecolor{myRed}{RGB}{195,10,10}
\definecolor{myGreen}{RGB}{55,149,73}

\newcommand\blfootnote[1]{%
  \begingroup
  \renewcommand\thefootnote{}\footnote{#1}%
  \addtocounter{footnote}{-1}%
  \endgroup
}

\usepackage{hyperref}
\hypersetup{colorlinks = True, 
	    linkcolor = red, 
        citecolor = blue} 
\begin{document}
\title{Promptable Counterfactual Diffusion Model for Unified Brain Tumor Segmentation and Generation with MRIs}
%
\author{
Yiqing Shen\textsuperscript{*} \and Guannan He\textsuperscript{*} \and Mathias Unberath\textsuperscript{(\Letter)}
}
\authorrunning{Y. Shen et al.}
\institute{
Johns Hopkins University, Baltimore, MD, 21218, USA\\
\email{\{yshen92,unberath\}@jhu.edu}
}

\maketitle   
\blfootnote{* Equal contributions. }
\begin{abstract}
Brain tumor analysis in Magnetic Resonance Imaging (MRI) is crucial for accurate diagnosis and treatment planning.
However, the task remains challenging due to the complexity and variability of tumor appearances, as well as the scarcity of labeled data. 
Traditional approaches often address tumor segmentation and image generation separately, limiting their effectiveness in capturing the intricate relationships between healthy and pathological tissue structures.
We introduce a novel promptable counterfactual diffusion model as a unified solution for brain tumor segmentation and generation in MRI. 
%
%
%
The key innovation lies in our mask-level prompting mechanism at the sampling stage, which enables guided generation and manipulation of specific healthy or unhealthy regions in MRI images.
Specifically, the model's architecture allows for bidirectional inference, which can segment tumors in existing images and generate realistic tumor structures in healthy brain scans.
Furthermore, we present a two-step approach for tumor generation and position transfer, showcasing the model's versatility in synthesizing realistic tumor structures. 
Experiments on the BRATS2021 dataset demonstrate that our method outperforms traditional counterfactual diffusion approaches\cite{counterfactual}, achieving a mean IoU of 0.653 and mean Dice score of 0.785 for tumor segmentation, outperforming the 0.344 and 0.475 of conventional counterfactual diffusion model. 
Our work contributes to improving brain tumor detection and segmentation accuracy, with potential implications for data augmentation and clinical decision support in neuro-oncology.
The code is available at \url{https://github.com/arcadelab/counterfactual_diffusion}.

\end{abstract}

\section{Introduction}
Brain tumor analysis in Magnetic Resonance Imaging (MRI) is an important task in medical diagnosis and treatment planning\cite{ce2023artificial}.
The accurate segmentation and characterization of tumors from MRI scans are essential for effective patient care in neuro-oncology\cite{khalighi2024artificial}.
However, this task remains challenging due to the complexity and variability of tumor appearances across different MRI modalities and the need for precise delineation of tumor boundaries.
Recent advancements in deep learning, particularly in the fields of computer vision and medical image analysis, have shown promising results in addressing these challenges by training an end-to-end model like UNet\cite{conze2023current}.
Diffusion models, which have demonstrated remarkable capabilities in image generation tasks, offer a new perspective on medical image analysis by unifying discriminative and generative tasks \cite{kazerouni2023diffusion}.
%

Despite these advancements, several key challenges persist in brain tumor analysis. High-quality, annotated MRI datasets for brain tumors are scarce, hindering the development of robust segmentation models. 
The diverse appearance of brain tumors across patients and MRI modalities makes it difficult for traditional segmentation models to generalize effectively. 
Many existing deep learning models operate as ``black boxes'', making it challenging for clinicians to understand and trust their decisions. 
Furthermore, current approaches often struggle to generate realistic tumor images or manipulate tumor characteristics in a controlled manner.
To address these challenges, we innovatively propose a unified approach that combines the strengths of diffusion models, counterfactual reasoning, and prompt-guided generation. 
This combination offers several advantages. 
Counterfactual reasoning allows the model to learn the relationship between healthy and pathological tissue, potentially improving segmentation accuracy and generalization. 
Prompt-guided generation enables more controlled and interpretable tumor analysis, allowing clinicians to interact with the model and explore various scenarios. 
The unified framework for both segmentation and generation can leverage the synergies between these tasks, potentially improving performance on both fronts. 
%
%
For segmentation, our approach aims to accurately identify and delineate tumor regions across multiple MRI modalities (T1, T1-ce, T2, and FLAIR). 
In terms of tumor generation, we focus on creating synthetic tumor images and achieving tumor position transfer, which has implications for data augmentation and hypothetical disease progression modeling.
%

%
%

The key contributions of our work are three-fold. 
Firstly, we present a promptable counterfactual diffusion sampling method that outperforms traditional approaches in both tumor segmentation and generation tasks by enabling manual intervention in the diffusion denoising process. 
Secondly, we provide a comprehensive comparison of Transformer-based and UNet-based denoising architectures in the context of the counterfactual diffusion model.
We show that Transformer can capture global context more effectively than traditional convolutional approaches, leading to improved performance in handling complex tumor morphologies.
We also introduce a two-step tumor generation approach, demonstrating the versatility of our method in manipulating and synthesizing realistic tumor structures. 
%

\begin{figure}[htbp!]
    \centering
    \includegraphics[width=\linewidth]{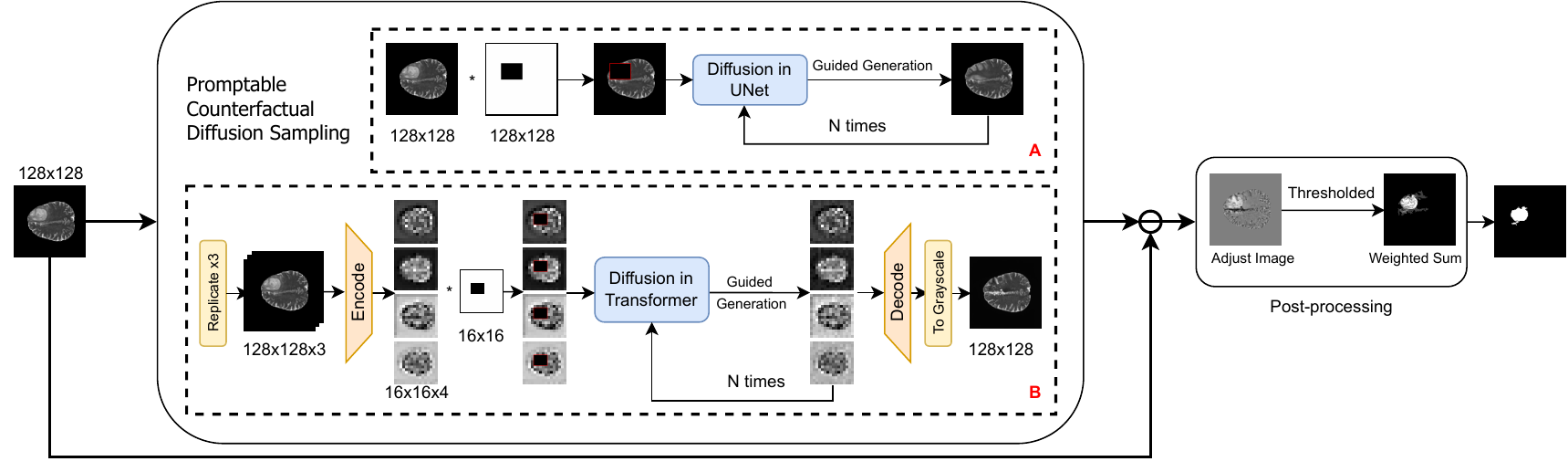}
    \caption{The overall pipeline of the promptable counterfactual diffusion sampling scheme. (A) UNet-based approach: A $128\times128$ input image is combined with a binary mask prompt for guided generation through a UNet-based denoising network, iterated N times. 
    (B) Transformer-based approach: The input image is replicated, encoded, and processed at multiple resolutions ($128\times128\times3$, $16\times16\times4$) before undergoing guided generation through a Transformer-based denoising network, also iterated N times. 
    Both approaches leverage mask-guided diffusion for region-specific generation. 
    The final output undergoes post-processing, including thresholding and weighted sum computation, to produce the segmented tumor region.}
    \label{fig:1}
\end{figure}

\section{Methods}
In this section, we detail the architecture of our promptable counterfactual diffusion model.
We further enhance the diffusion process for the counterfactual diffusion model \cite{counterfactual} by replacing the conventional U-Net denoising network \cite{ho2020denoising} backbone with a Transformer-based denoising network \cite{dit2023} to show its improvement is also applicable beyond DDPM.
The overall framework is summarized in Fig.~\ref{fig:1}.

\subsubsection{Promptable Counterfactual Diffusion Sampling}
Our proposed promptable counterfactual diffusion sampling scheme builds upon the DDPM \cite{ho2020denoising}.
It enables both prompt-guided tumor segmentation \cite{amit2021segdiff} and generation within a unified diffusion model by formulating them uniformly as a counterfactual diffusion process. 
%
%
%
Inspired by previous work \cite{lugmayr2022repaint}, our promptable counterfactual diffusion sampling scheme extends this approach to allow for the generation of specific areas in MRI images as either healthy or unhealthy, based on a given mask prompt.
During the reverse diffusion process, we incorporate mask prompts to guide the generation of desired regions, ensuring semantic consistency with the rest of the MR image.
By integrating this prompting mechanism with DDPM's iterative denoising process, we can produce high-quality, diverse counterfactual outputs for both healthy and tumor regions.
This flexibility enables the use of a single model for both segmentation and augmentation in MRI analysis, streamlining the workflow and potentially improving overall performance.
The complete process of our promptable counterfactual diffusion sampling scheme is detailed in Alg.~\ref{tab:prompt_algro} and illustrated.

\begin{algorithm}[htbp]
\caption{Promptable Counterfactual Diffusion Sampling Scheme}\label{tab:prompt_algro}
\begin{algorithmic}[1]
\Require Clean image $x_0$, number of timesteps $T$, mask prompt $m$, guidance signal $c$, neural network $\epsilon_\theta$
\Ensure Generated counterfactual image $x_{0}^{\mathrm{gen}}$

\State Initialize $x_T \sim \mathcal{N}(0, \mathbf{I})$

\For{$t = T$ to $1$}
    \State \textbf{Sample Known Regions:}
    \State $x_{t-1}^{\mathrm{known}} \sim q(x_{t-1} | x_t, x_0)$ where
    \State $q(x_{t-1} | x_t, x_0) = \mathcal{N}(\tilde{\mu}_t(x_t, x_0), \tilde{\beta}_t \mathbf{I})$
    \State $\tilde{\mu}_t(x_t, x_0) = \frac{\sqrt{\bar{\alpha}_{t-1}}\beta_t}{1-\bar{\alpha}_t}x_0 + \frac{\sqrt{\alpha_t}(1-\bar{\alpha}_{t-1})}{1-\bar{\alpha}_t}x_t$
    
    \State \textbf{Guided Sampling for Unknown Regions:}
    \State Predict noise: $\hat{\epsilon} = \epsilon_\theta(x_t, t)$
    \State Compute denoised estimate: $x_0^t = \frac{x_t - \sqrt{1-\bar{\alpha}_t}\hat{\epsilon}}{\sqrt{\bar{\alpha}_t}}$
    \State Apply guidance: $x_0^t \leftarrow x_0^t + c \cdot \nabla_{x_0^t} \log p(x_0^t)$
    \State $x_{t-1}^{\mathrm{guided}} \sim \mathcal{N}(\mu_\theta(x_t, t, x_0^t), \Sigma_\theta(x_t, t))$
    
    \State \textbf{Combine Samples:}
    \State $x_{t-1} = m \odot x_{t-1}^{\mathrm{known}} + (1 - m) \odot x_{t-1}^{\mathrm{guided}}$
\EndFor

\State \textbf{Output:} $x_{0}^{\mathrm{gen}} \gets x_{0}$
\end{algorithmic}
\end{algorithm}

\subsubsection{Transformer-based Denoising Network}
We further adopt the Diffusion Transformer (DiT) \cite{dit2023} for the promptable counterfactual diffusion sampling scheme in brain tumor segmentation and generation from MRI. 
Specifically, our approach replaces the traditional U-Net denoising network \cite{unet} with a Transformer-based backbone, leveraging the superior global context capture capabilities of Transformers \cite{vaswani2017attention} to enhance the diffusion model's performance in generating high-quality MRI images.
Building on the foundational concepts of DiT, we introduce several key enhancements tailored to MRI. 
Firstly, we integrate a prompting mechanism for guided tumor segmentation and generation, enabling the creation of specific healthy or unhealthy regions in MRI images based on mask inputs and user-defined prompts.
This exploits the Transformers' global context awareness to achieve more precise and contextually relevant image generation.
Second, the preprocessing of input MRI data involves utilizing a pre-trained variational autoencoder (VAE) \cite{kingma2013auto} from Stable Diffusion \cite{stablediff} to encode input images into a latent space. 
These latent representations are then transformed into token sequences through patchification. 
Our processing pipeline processes these token sequences through multiple DiT blocks. 
Each DiT block incorporates in-context conditioning, cross-attention, and adaptive layer normalization (adaLN), similar to adaptive normalization layers in GANs \cite{perez2018film}. 
These components collectively improve the denoising network's ability to manage conditioning information, enhancing both performance and scalability.
Finally, in this work, we implement the DiT B-2 architecture size, as a balance between computational efficiency and model performance. 
Finally, to address the RGB input requirement of the pre-trained VAE, we process each MRI modality separately by replicating it across three channels, creating an RGB-like input.

\subsubsection{Initial Segmentation Derivation}
The initial segmentation is obtained by comparing the original MRI image (factual) with the generated healthy version (counterfactual) produced by our model. 
Specifically, we compute the difference image between these two:
\begin{equation}
D(x, y) = |I_f(x, y) - I_c(x, y)|,
\end{equation}
where $I_f(x, y)$ is the intensity of the factual image at location $(x, y)$, and $I_c(x, y)$ is the intensity of the counterfactual image at the same location. 
This difference image $D(x, y)$ highlights areas where the model has made changes, potentially indicating tumor regions.
The resulting difference image serves as our preliminary segmentation.
However, this raw output may contain noise or ambiguities, necessitating further refinement through post-processing.

\subsubsection{Post-Processing Refinement for Segmentation}
After obtaining preliminary tumor segmentation results using the proposed promptable counterfactual diffusion sampling scheme, we apply a series of post-processing techniques to refine these results, ensuring higher accuracy and robustness. 
Our post-processing pipeline consists of three main steps: contrast and brightness adjustment, thresholding, and channel-weighted sum computation.

First, we enhance the initial segmentation output, which represents the difference image between the ground truth and the generated healthy image, by adjusting its contrast and brightness. 
This step amplifies the contrast and highlights tumor regions more clearly, facilitating easier distinction.
Next, we convert the contrast-adjusted image into a binary image using Otsu's thresholding method \cite{Otsu} from the \texttt{skimage.filters} module. 
It determines the optimal threshold value to effectively separate tumor regions from the background.
Finally, we process each of the four channels corresponding to the four MRI modalities separately. 
We assign an equal weight of 0.25 to each modality's result and integrate them by summing the weighted outputs. 
A voxel is classified as having a tumor if the cumulative weight is 0.5 or more, ensuring that tumor presence is determined based on majority overlap across channels. 
This process is formalized as
\begin{equation}
S(x, y) = \sum_{i=1}^{4} w_i \cdot M_i(x, y)
\label{eq:weighted_sum}
\end{equation}
where $w_i = 0.25$  for each modality $i$, and $M_i(x, y)$ represents the binary mask for modality $i$ at location $(x, y)$. 
A tumor is identified at location $(x, y)$ if $S(x, y) \geq 0.5$. 
It ensures that a location is classified as a tumor if half or more of the modalities indicate the presence of a tumor.

\section{Results}

\subsubsection{Datasets}
Our model training and experiments utilized the BRATS2021 (Brain Tumor Segmentation Challenge 2021) dataset \cite{brats1,brats2,brats3}, a well-established MRI benchmark for brain tumor segmentation tasks. 
This dataset comprises multimodal MRI scans from patients, featuring four MRI sequences per patient: T1-weighted (T1), post-contrast T1-weighted (T1ce), T2-weighted (T2), and Fluid Attenuated Inversion Recovery (FLAIR). 
%

\subsubsection{Implementation Details}
We employed data from 1,000 patients for training, with each patient contributing 155 2D slices per modality, where we randomly split 104 slices as the test set.
All implementations are made on the 2D slice level.
For each slice, we concatenated the four modalities along the channel dimension into a single file. These images were then resized to $128\times128$ pixels and normalized. 
In our training process, we classified axial slices containing at least one tumor pixel as ``unhealthy,'' while those without any tumor pixels were considered ``healthy''. These two categories of data were used to train the conditional diffusion model. 
We conducted the training process on a single NVIDIA RTX A6000 48 GB GPU, optimizing our model with a learning rate of 1e-4, the AdamW optimizer, a batch size of 128, and training for 15,000 steps.
We assessed the performance of tumor segmentation using two primary metrics: (i) the Dice coefficient and (ii) Intersection over Union (IoU).
%
%
%
To evaluate our promptable counterfactual diffusion sampling method, we compared it with the traditional counterfactual sampling method\cite{counterfactual}. We assessed four methods in total:
(i) Promptable counterfactual diffusion sampling with a Transformer denoising network (our proposal), 
(ii) Promptable counterfactual diffusion sampling with a UNet denoising network (our proposed method, but as ablation), (iii) Counterfactual sampling with a Transformer denoising network (\textit{i}.\textit{e}., without mask prompt, baseline), and (iv) Counterfactual sampling with a UNet denoising network (\textit{i}.\textit{e}., without mask prompt, baseline). 
%
%

\begin{table}[t!]
\centering
\caption{Comparison of brain tumor segmentation performance on the BRATS2021 dataset.
We compare against promptable vs. traditional counterfactual diffusion model\cite{counterfactual} with Transformer and UNet as denoising networks.}
\label{tab1}
\begin{tabular}{lcc}
\toprule
Method & Mean IoU & Mean Dice \\
\midrule
Promptable Counterfactual w/ Transformer & \textbf{0.653} & \textbf{0.785} \\
Promptable Counterfactual w/ UNet & 0.647 & 0.772 \\
Counterfactual Diffusion w/ Transformer & 0.366 & 0.479 \\
Counterfactual Diffusion w/ UNet & 0.344 & 0.475 \\
\bottomrule
\end{tabular}
\end{table}

\begin{figure}[htbp]
    \centering
    \includegraphics[width=0.65\linewidth]{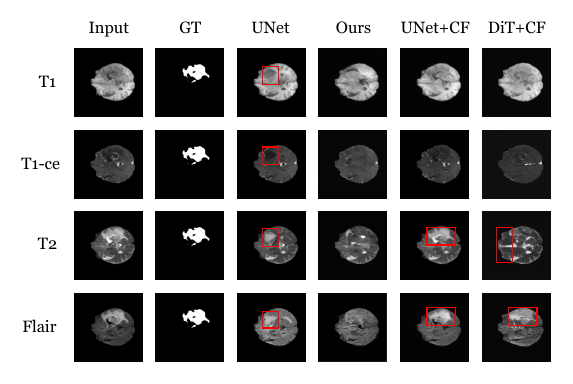}
    \caption{Comparative visualization of brain tumor segmentation results across four MRI modalities (T1, T1-ce, T2, and Flair). The columns show: Input (original MRI), GT (ground truth), UNet (our PCF+UNet), Ours (our PCF+Transformer), UNet+CF (Counterfactual Diffusion with UNet), and DiT+CF (Counterfactual Diffusion with Transformer). Red boxes highlight areas where methods struggled to accurately segment or remove tumors. Our PCF+Transformer method (Ours) demonstrates superior tumor removal across all modalities.}
    \label{fig:repaint_example1}
\end{figure}

\subsubsection{Evaluation of Tumor Segmentation}
Our assessment encompassed four distinct methods: Promptable Counterfactual with Transformer (PCF+Transformer), Promptable Counterfactual with UNet (PCF+UNet), Counterfactual Diffusion Sampling with Transformer (CF+Transformer), and Counterfactual Diffusion Sampling with UNet (CF+UNet) \cite{counterfactual}. 
The quantitative results of this evaluation are presented in Table \ref{tab1}.
Our proposed method (\textit{i}.\textit{e}., PCF+Transformer) demonstrated superior performance, achieving the highest Mean IoU of 0.653 and Mean Dice score of 0.785. 
This outstanding performance underscores the effectiveness of our approach in accurate tumor segmentation. 
Notably, both promptable counterfactual methods (PCF+Transformer and PCF+UNet) consistently outperformed their traditional counterfactual diffusion\cite{counterfactual} counterparts, highlighting the significant advantages of guided generation in medical imaging tasks.
%
%
The visual examples of our segmentation results, as illustrated in Fig.~\ref{fig:repaint_example1}, provide further insight into the performance of each method.
PCF+Transformer exhibited the most impressive results, effectively identifying and delineating tumor regions across all MRI modalities. 
While PCF+UNet also showed strong performance, it occasionally fell short of completely delineating tumor boundaries.
In contrast, the CF methods demonstrated inconsistent performance across different MRI modalities, with some instances of incomplete tumor identification or false positives.
These results collectively emphasize the superiority of our promptable counterfactual approach, particularly when coupled with a Transformer denoising network.

\begin{figure}[t!]
    \centering
    \includegraphics[width=0.65\linewidth]{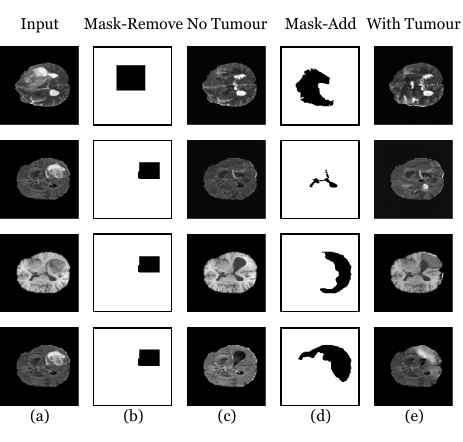}
    \caption{Visualization of the tumor regeneration process: (a) Original unhealthy slice, (b) Mask for removing the tumor, (c) Healthy slice after tumor removal, (d) Random mask for tumor generation (e) Regenerated tumor in a new location.}
    \label{fig:uhu}
\end{figure}

\subsubsection{Evaluation of Tumor Generation}
We developed a two-step approach to generate tumor images and achieve tumor position transfer, leveraging the capabilities of our promptable counterfactual model.
The first step is tumor removal, where we utilize our promptable counterfactual model to transform unhealthy slices containing tumors into healthy slices. 
This process involved the precise erasure of tumor regions while preserving the integrity of surrounding brain tissues. 
It demonstrates the model's ability to comprehend and manipulate complex anatomical structures within MRI images.
The second step is tumor regeneration, which employs randomly generated masks to guide the regeneration of tumors within the previously ``healed'' slices. 
By placing these masks, we were able to generate new tumor regions in different locations, effectively achieving tumor position transfer. 
This capability showcases the model's flexibility in synthesizing realistic tumor structures while maintaining anatomical plausibility.
The results of this two-step process are visually demonstrated in Fig.~\ref{fig:uhu}. This figure illustrates the progression from original tumor-containing images, through the tumor removal phase, to the final images with regenerated tumors in new positions.
This approach not only demonstrates the versatility of our promptable counterfactual model but also opens up new possibilities for data augmentation.
By generating diverse tumor presentations, we can potentially provide valuable training data for medical professionals. 
%

\section{Conclusion}
In this paper, we introduced a promptable counterfactual diffusion sampling method, which demonstrates the unification of tumor segmentation and generation tasks.
Our experiments revealed important distinctions between the UNet and DiT denoising network backbones. 
The UNet-based model, while effective in many scenarios, occasionally generated black regions when presented with large masks or those extending beyond the brain shape in the MRI. 
In contrast, the DiT backbone, operating in the latent space, demonstrated superior robustness and stability in handling these complex cases.
Despite the lower resolution resulting from latent space processing, the DiT backbone excelled in managing intricate masks more effectively than its UNet counterpart.
%
%
%
Our approach can remove the tumor while maintaining the integrity of the original image structure, representing a significant improvement in the field.
%
Future work could explore its application to other medical imaging domains and investigate its potential in clinical decision support systems. 
\bibliographystyle{splncs04}
\bibliography{main.bib}

\end{document}